# Combinations of Jaccard with Numerical Measures for Collaborative Filtering Enhancement: Current Work and Future Proposal

Ali A. Amer[1], Loc Nguyen[2]

[1]Computer Science Department, Taiz University, Yemen, aliaaa2004@yahoo.com
[2]Loc Nguyen's Academic Network, Vietnam, Email: ng_phloc@yahoo.com

**Abstract**
Collaborative filtering (CF) is an important approach for recommendation system which is widely used in a great number of aspects of our life, heavily in the online-based commercial systems. One popular algorithms in CF is the K-nearest neighbors (KNN) algorithm, in which the similarity measures are used to determine nearest neighbors of a user, and thus to quantify the dependency degree between the relative user/item pair. Consequently, CF approach is not just sensitive to the similarity measure, yet it is completely contingent on selection of that measure. While Jaccard - as one of those commonly used similarity measures for CF tasks - concerns the existence of ratings, other numerical measures such as cosine and Pearson concern the magnitude of ratings. Particularly speaking, Jaccard is not a dominant measure, but it is long proven to be an important factor to improve any measure. Therefore, in our continuous efforts to find the most effective similarity measures for CF, this research focuses on proposing new similarity measure via combining Jaccard with several numerical measures. The combined measures would take the advantages of both existence and magnitude. Experimental results on, Movie-lens dataset, showed that the combined measures are preeminent outperforming all single measures over the considered evaluation metrics.

**Keywords:** Collaborating Filtering, Similarity Measure, Jaccard, K-Nearest Neighbors algorithm, Recommendation Systems.

## 1. Introduction

With the explosive increase of digital data, Recommendation System (RS) is being highly necessary to filter the constantly-accumulated flow of information and data online. RS should has the ability to offer the desired recommendations referring to the user preferences [1]. Recommendation system is a system which recommends items to users among many existing items in the database. In its turn, item is anything which users consider, such as product, book, and newspaper. On the whole, there are three main approaches for recommendation: content-based filtering (CBF), collaborative filtering (CF), and hybrid-based techniques [2]. CBF exploits the user's information (gender, age, etc.) to predict his preferences without considering others users' information. On the other hand, CF recommends the item to any user if her/his neighbors are interested in the same item. Generally speaking, CF-based RS has two models: memory-based and model-based methods [3]. The memory-based method utilizes the entire database to discover a set of users/items similar to the target user/item. While the user-based method seeks to construct the model for user behavior to forecast his selections. On one hand, the memory-based method usually provide higher precision while the model-based method is more efficient on large data sets. On the other hand, the model-based methods are more complex as they include training a model and tuning several hyperparameters [4].

CF-based RS often work by comparing the new case with previous cases. It studies the behavior of a set of users who are analogous to the active user, or the characteristics of a set of items which are similar to the target item. In general, the more the users share common items, the higher the similarity is between users. On the whole, there two key methods of CF: User-based CF and Item-based CF. In user-base CF, via user x user matrix, the flow of incoming items is selected depending on the evaluation provided by other users who have already liked the items. So, if item has been defined interesting by the relative user, it would automatically be suggested to users who have similar opinions in common. On the other hand, in item-based model, via item x item matrix, RS will suggest items which are highly similar to the already-rated items with a maximal score by the active "target" user. Then, the predicted rating based on the similarity degree between the item and its neighbor. That is, the higher the similarity is, the more the predicted rating is analogues to the rating's neighbors.

One popular algorithm in CF, on the other hand, is the K-nearest neighbors (KNN) algorithm. KNN algorithm [5] aims to find out nearest neighbors of the regarded user (known as active user), and then to recommend active user's items which are liked by user's neighbors. Thus, the essence of KNN algorithm is to use similarity measures to discover the nearest neighbors of the active rating vector [6]. So, this research is directed to find effective similarity measures for CF. Among the most popular similarity measures are cosine, Pearson and Jaccard. Owing to the importance of similarity measure in CF process, selecting the best-fit similarity measure among a good number of measures proposed in literature is being intelligibly challenging task. Consequently, in this work, in our ongoing efforts, we aim to find highly-effective similarity measures via the combination of Jaccard with numerical measures like cosine and PCC. The main contribution of this research paper is represented in proposing new similarity measures via combining the jaccard measure with numerical measures. The key aim is to experimentally prove that jaccard measure can effectively contribute in enhancing the performance of numerical measures.

## 2. Literature Review
In CF literature, dozens of similarity measures and machine learning models are separately applied on CF-based recommendation Systems (RS), and their effects recorded either in user-based or item-based models [2, 7-8]. For instance, [9-12] utilized the contextual information of users to present similarity measures based on the singularity factor. While [11] proposed Context Based Rating Prediction (CBRP) to find the context score of each nominated user for the considered pair "user-item". In [9] the Mean-Jaccard-Differences, MJD, was presented to improve the traditional similarity measures using the singularity of user ratings. In [13], several traditional similarity measures (PCC, cosine and some distance metrics) were combined to introduce a combined similarity measure. A similarity measure called (IPWR), short for improved PCC weighted with user rating preference behavior (RPB), was presented in [14] by combining PCC with RPB. IPWR was seen superior over the state-of-art measures. PIWR was shown better than traditional measures. Similarly, [15-16] produced new measures to select neighbors based on the neighborhood union and intersection. Like jaccard, these measures were reliant on the shared items when finding the neighbors of user of interest. The similarity between items would have zero value when there were non-shared items between the intended users, making these measures faulty in this case. On the other extreme, using Bhattacharyya coefficient, several studies have been dedicated to solve the dilemma of data sparsity [17-19]. Meanwhile, [20] presented subspace clustering-driven measure to tackle both problems of the high dimensionality and the data sparsity. A fast neighbor user searching (FNUS) approach was proposed in [21] to enhance RS performance via generating the item subspaces into interested item, neither interested nor uninterested (NINU) item, and uninterested item subspaces. Then, the co-rated item numbers between a target user and other users were computed, and used to find the three subsets of neighbor users for the target user. Through the union of the three neighbor user subsets, the final neighbor user set is drawn. The FNUS was seen having a competitive rendering on huge datasets.

In [22], a linear combination was proposed. Using PSS, Bhattacharya Coefficient, and Jaccard, the preferences and local context of users' behavior and the percentage of shared ratings between each user pair were considered. In [23-24], new measures, based on the global user preference and context information, were proposed to enhance RS accuracy. In [25], a Wasserstein Collaborative Filtering (WCF) technique was proposed. Under user embedding constraint, WCF predicted user preference on cold-start items by minimizing the Wasserstein distance. Ahn [26] investigated the deficits of traditional similarity measures in CF. Using the specific meanings of co-ratings and the explanation of user ratings, author then introduced a heuristic similarity measure called PIP which stands for three semantic heuristics, namely, Proximity, Impact and Popularity. In [27], however, PIP was shown faulty in some cases of data sparsity. So, a PIP-based heuristic similarity model called (NHSM) was introduced to tackle the limitation of PIP effectively.

On the same page, in [28], the data sparsity was demonstrated to have a devastating impact on the performance of recommendation systems. Even though the PIP and NHSM measures [26] provided an improved solution for sparsity problem, the range of values for each component in PIP is very high. So, a modified proximity-impact-popularity (MPIP) similarity measure was introduced in [28]. The MPIP expression was designed to close the Gab of PIP measure whose range of values for each component was very high. MPIP was shown better than PIP and other competitive measures. In [6], using cross validation, authors designed three efficient similarity measures to effectively tackle the data sparsity problem, namely, difference-based similarity measure (SMD), hybrid difference-based similarity measure (HSMD), and triangle-based cosine measure (TA). SMD and TA were proven to be superior comparing with their rivals. In [4], the latent semantic integrated explicit rating (LSIER) scheme was presented to enhance RS performance. The LSIER scheme was designed by combining the probabilistic latent semantic index (PLSI) model which was used to train user's access records, and the probabilistic matrix factorization (PMF) model which was used to give the user feature and service feature matrices. Finally, using the domain sensitivity, in [29], a sentiment-based model with contextual information was developed for RS,

Finally, in [30], a new similarity measure was presented to fully use the limited rating information of cold users. As first metric, the Popularity-Mean Squared Difference metric was designed to take the influence of popular items into account, the difference between each user pair was averaged in terms of common ratings and non-numerical information of ratings. Second, the Singularity-Difference was designed to hold the deviation degree of favor to items between user pair. Experiment results, on MovieLens, Epinions and Netflix, showed that the designed measure outperformed seven popular similarity measures in terms of MAE, precision, recall and F1-Measure.

In parallel with the presented-above measures, our proposed future project will come to meticulously address all the drawn-above limitations of CF. The cold start "data sparsity" and high dimension problems will be thoroughly analyzed and carefully tackled using the proposed measures which would also consider the contextual information of users, the dilemma of high dimensionality, and implicit user's feedback. Most importantly, a comprehensive framework for CF similarity measures, that would include an almost "60-90" measures, is set to be introduced. Moreover, to promote performance of CF, new variations of KNN, CNN, SVM and MNB will be developed, and later merged with top-performer similarity measures.

### 2.1. The Used Similarity Measures

Given two rating vectors $u_1 = (r_{11}, r_{12},…, r_{1n})$ and $u_2 = (r_{21}, r_{22},…, r_{2n})$ of user 1 and user 2, in which user 1 is an active user and some $r_{ij}$ can be missing (they are empty). Let $I_1$ and $I_2$ be set of indices of items that user 1 and user 2 rated, respectively. Let $I = I_1 \cap I_2$ denote intersection set of $I_1$ and $I_2$ and let $I_1 \cup I_2$ denotes union set of $I_1$ and $I_2$. Notation |x| indicates absolute value of number, length of vector, length of geometric segment, or cardinality of set, which depends on context. Let sim($u_1$, $u_2$) denote the similarity of $u_1$ and $u_2$. For instance, the cosine measure of $u_1$ and $u_2$ is defined as follows:

$$\text{sim}(u_1, u_2) = \text{cosine}(u_1, u_2) = \frac{\sum_{j \in I_1 \cap I_2} r_{1j} r_{2j}}{\sqrt{\sum_{j \in I_1 \cap I_2} (r_{1j})^2} \sqrt{\sum_{j \in I_1 \cap I_2} (r_{2j})^2}} \quad (1)$$

Cosine measure is effective, yet it has a drawback that in the dimensional space, there may be two end points of two vectors which are far from each other according to Euclidean distance, but their cosine is high. This is negative effect of Euclidean distance which decreases the accuracy of cosine similarity. Therefore, as an advanced cosine, the triangle area (TA) measure [6] was proposed as an improved version of cosine measure. Fig. 1 illustrates TA measure.

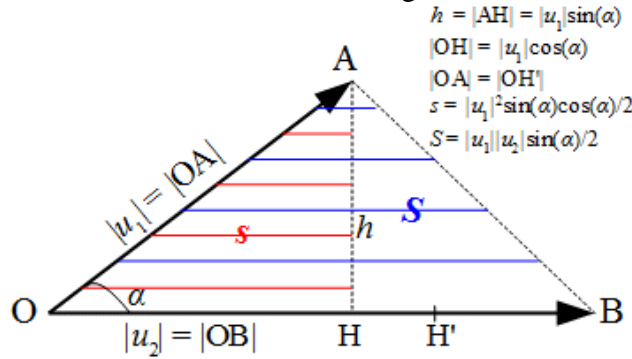

**Figure 1.** Triangle area (TA) measure with $0 \leq \alpha \leq \pi/2$

TA measure uses the ratio of basic triangle area to the whole triangle area as reinforced factor for Euclidean distance so that it can alleviate negative effect of Euclidean distance, whereas it keeps simplicity and effectiveness of both cosine measure and Euclidean distance. TA is defined as follows:

$$u_1 \cdot u_2 \geq 0: \text{TA}(u_1, u_2) = \begin{cases} \frac{(u_1 \cdot u_2)^2}{|u_1|(|u_2|)^3} \text{ if } |u_1| \leq |u_2| \\ \frac{(u_1 \cdot u_2)^2}{(|u_1|)^3|u_2|} \text{ if } |u_1| > |u_2| \end{cases}$$

$$u_1 \cdot u_2 < 0: \text{TA}(u_1, u_2) = \begin{cases} \frac{u_1 \cdot u_2}{(|u_2|)^2} \text{ if } |u_1| \leq |u_2| \\ \frac{u_1 \cdot u_2}{(|u_1|)^2} \text{ if } |u_1| > |u_2| \end{cases} \quad (7)$$

Where $|u_1|$ and $|u_2|$ are lengths of $u_1$ and $u_2$, respectively whereas $u_1 \cdot u_2$ is the dot product (scalar product) of $u_1$ and $u_2$, respectively.

On the other hand, Pearson correlation is another popular similarity measure, which is defined as follows:

$$\text{Pearson}(u_1, u_2) = \frac{\sum_{j \in I_1 \cap I_2}(r_{1j} - \bar{u}_1)(r_{2j} - \bar{u}_2)}{\sqrt{\sum_{j \in I_1 \cap I_2}(r_{1j} - \bar{u}_1)^2} \sqrt{\sum_{j \in I_1 \cap I_2}(r_{2j} - \bar{u}_2)^2}} \quad (2)$$

Where $\bar{u}_1$ and $\bar{u}_2$ are mean values of $u_1$ and $u_2$, respectively.

$$\bar{u}_1 = \frac{1}{|I_1|} \sum_{j \in I_1} r_{1j}, \bar{u}_2 = \frac{1}{|I_2|} \sum_{j \in I_2} r_{2j}$$

Liu et al., (2013) proposed a new similarity measure called NHMS to improve recommendation task in which only few ratings are available. Their NHMS measure is contingent on the sigmoid function and the improved PIP measure as PSS (*Proximity – Significance – Singularity*). PSS similarity is calculated as follows:

$$\text{PSS}(u_1, u_2) = \sum_{j \in I} \text{Proximity}(r_{1j}, r_{2j}) * \text{Significance}(r_{1j}, r_{2j}) * \text{Singularity}(r_{1j}, r_{2j}) \quad (3)$$

Followings are the equations of factors such as proximity, significance, and singularity based on sigmoid function.

$$\text{Proximity}(r_{1j}, r_{2j}) = 1 - \frac{1}{1 + \exp(-|r_{1j} - r_{2j}|)}$$

$$\text{Significance}(r_{1j}, r_{2j}) = \frac{1}{1 + \exp(-|r_{1j} - r_m||r_{2j} - r_m|)}$$

$$\text{Singularity}(r_{1j}, r_{2j}) = 1 - \frac{1}{1 + \exp\left(-\left|\frac{r_{1j} + r_{2j}}{2} - \mu_j\right|\right)}$$

Note, $r_m$ be median of rating values.

Meanwhile, Jaccard measure is ratio of cardinality of common set $I_1 \cap I_2$ to cardinality of union set $I_1 \cup I_2$. It measures how much common items both users rated, which is defined as follows:

$$\text{Jaccard}(u_1, u_2) = \frac{|I_1 \cap I_2|}{|I_1 \cup I_2|} \quad (4)$$

Liang et al., (2015) proposed an improved variant of Jaccard based on the concept of singularity. Their measure IJ is specified as follows:

$$\text{IJ}(u_1, u_2) = \left( \sum_{j \in PA} S_P^j + \sum_{j \in NA} S_N^j + \sum_{j \in D} \sqrt{S_P^j S_N^j} \right)$$
$$/ \left( \sum_{j \in PA} S_P^j + \sum_{j \in NA} S_N^j + \sum_{j \in D} \sqrt{S_P^j S_N^j} + \sum_{j \in PO} \sqrt{S_P^j S_E^j} + + \sum_{j \in NO} \sqrt{S_N^j S_E^j} \right) \quad (5)$$

Where $S_j^P$, $S_j^N$, and $S_j^E$ are positive singularity, negative singularity, and empty singularity of item j, respectively. The *PA*, *NA*, *D*, *PO*, and *NO* are sets of agreed positive ratings, agreed negative ratings, disagreed ratings, positive ratings, and negative ratings, respectively.

On the other hand, Jaccard could be combined with any measure. For example, CosineJ is combination of Jaccard, and cosine is formulated as follows:

$$\text{CosineJ}(u_1, u_2) = \text{cosine}(u_1, u_2) * \text{Jaccard}(u_1, u_2) = \frac{\sum_{j \in I_1 \cap I_2} r_{1j} r_{2j}}{\sqrt{\sum_{j \in I_1}(r_{1j})^2} \sqrt{\sum_{j \in I_2}(r_{2j})^2}} * \frac{|I_1 \cap I_2|}{|I_1 \cup I_2|} \quad (6)$$

In general, cosine, Pearson, and PSS are numerical measures because they are calculated based on real rating values, but they cannot solve the problem of missing values. At the other hand, Jaccard only focuses on the existence of rating values but it ignores the magnitude of ratings. The combination of Jaccard and numerical measure would take the advantages of both existence and magnitude, which is described in methodology section.

The rest of this article goes as follows. Section 2 scan the most relevant CF works. In section3, the methodology is concisely described. Section 4 provided constructive discussion over the findings, and the conclusion and future works are given in section 5.

## 3. Methodology

### 3.1. Motivations
In collaborative filtering (CF), offering favorites to users is essentially depending on the analysis of users' preferences and the existing correlation between their preferences. Most CF analysis use K-nearest neighbors (KNN) algorithm which is greatly affected by similarity measures and dataset sparseness. According to CF literature, some frequently used traditional measures like Cosine and PCC do not reach the desired performance as they place a complete emphasize on the co-related ratings and disregard the non-co-rated items when users' correlations are investigated. Although acceptable accuracy, they fail to effectively address the data sparsity problem (cold-start problem). Consequently, a few studies have been proposed to solve this problem. Nevertheless, these studies still suffer from data sparsity problem [1, 12]. Moreover, most of these studies evade testing similarity measures on item-based model due to the task complexity. Finally, till recently, there has been no efficient solution benchmarked to be universal and stable for CF performance. On the other hand, besides the way in which KNN is being designed, the key problem of KNN is the similarity computation when CF performance is affected by dataset's sparsity where most items are not rated.

### 3.2. Proposed Method
As a matter of fact, Jaccard improves accuracy of similarity of rating vectors because rating dataset has a lot of missing values whereas other measures depend only on existence of ratings. Put it different, Jaccard concerns both the existence and inexistence of ratings when missing values in incomplete rating dataset imply inexistence of ratings. According to our experimental results, albeit it does not consider the magnitude of ratings (real numbers), the more missing values the dataset has, the more accurate the Jaccard is. As a result, if rating dataset has enough rating values, Jaccard will be less accurate than other numerical measures. Concisely, Jaccard is not a dominant measure, but it is an important factor to improve any measure. Therefore, in this research, we combine Jaccard measures and other numerical measures in order to taking advantages of both existence and quantity of rating values. The combination is mutual, and not resonant and hence, we use multiplicative combination of Jaccard and other numerical measures such as cosine, Pearson, PSS, and TA. These numerical measures are typical with many variants. Furthermore, the advanced version of Jaccard which is developed by [30] is also combined with cosine, Pearson, PSS, and TA in comparison with Jaccard. This advanced measure is denoted by IJ in Table 1, which shows the combined measures used in this research.

**Table 1.** Combination of Jaccard and numerical measures

| Combined | Jaccard | Numerical measure | Formula |
|---|---|---|---|
| CosineJ | Jaccard | Cosine | Jaccard * cosine |
| PearsonJ | Jaccard | Pearson | Jaccard * Pearson |
| PSSJ | Jaccard | PSS | Jaccard * Pearson |
| TAJ | Jaccard | TA | Jaccard * TA |
| CosineIJ | IJ | Cosine | IJ * cosine |
| PearsonIJ | IJ | Pearson | IJ * Pearson |
| PSSIJ | IJ | PSS | IJ * Pearson |
| TAIJ | IJ | TA | IJ * TA |

The 8$^{th}$ combined measures such as CosineJ, PearsonJ, PSSJ, TAJ, cosineIJ, PearsonIJ, PSSIJ, and TAIJ are tested and compared with single and numerical measures cosine, Pearson, PSS, and TA. It is necessary to describe TA measure here. The next section mentions the experimental design and results.

## 4. Results

Dataset Movielens[1] is used for the experimental evaluation. It has 100,000 ratings from 943 users on 1682 movies (items), and every rating ranges from 1 to 5. In the experiments, dataset Movielens is divided into 5 folders and each folder includes training set and testing set. Training set and testing set in the same folder are disjoint sets. The ratio of testing set over the whole dataset depends on the testing parameter $r$. For instance, if $r = 0.1$, the testing set covers 10% of the dataset, which means that the testing set has 10,000 = 10%*100,000 ratings, and of course, the training set has 90,000 ratings. In the experimental design, parameter $r$ has nine values 0.1, 0.2, 0.3, 0.4, 0.5, 0.6, 0.7, 0.8, and 0.9. The smaller $r$ is, the more accurate measures are because training set gets large if $r$ gets small noting that KNN algorithm is executed on training set. Popular metrics to assess CF algorithms are mean absolute error (MAE), recall, and precision. MAE indicates accuracy of measures. The smaller MAE is, the more accurate the measures are, and so the better the algorithm is. Precision and recall are quality metrics that measure quality of recommended list – how much the recommendation list reflects user's preferences. The larger quality metric is, the better the algorithm is [33].

Quality of CF algorithm like KNN algorithm depends on both estimation and recommendation. Estimation is the ability to estimate or predict the exactly-missing values. Recommendation is the ability to provide the desired list of recommended items which should be as suitable as possible to users. Hence, different metrics (MAE, recall, precision) are used for different evaluation processes (estimation and recommendation). This independent evaluation allows us to test measures more objectively, in which the estimation process is directed on CF accuracy, and the recommendation process is directed on CF quality. In general, MAE is used for estimation whereas recall and precision are used for recommendation process.

It is also worth referring that the problem in recommendation is how to specify the number of recommended items (known as recommendation count) which is the length of recommended vector. We propose a technique to compute the recommendation count based on sparse-relevant ratio. Sparse-relevant ratio denoted $sr$ is specified as follows:

$$sr = \text{the-count-of-relevant-ratings} / (|\mathbf{U}| * |\mathbf{V}|)$$

---

[1] http://grouplens.org/datasets/movielens

Note, $|U|$ is the number of users and $|V|$ is the number of items. We then calculate recommendation count dynamically according to both dataset and each rating vector $u_i$. Let $C(u_i)$ be the recommendation count for user $i$, which means that KNN algorithms will recommend at least $C(u_i)$ items to user $i$. The recommendation count $C(u_i)$ is specified as follows:

$$C(u_i) = sr * (T - |I_i|) \qquad (8)$$

Where $T$ is the number of items noting that every item included in $T$ is rated by at least one user.

Table 2 shows MAE metric of all tested measures over all $r$ = 0.1, 0.2, 0.3, 0.4, 0.5, 0.6, 0.7, 0.8, and 0.9 within the estimation process. The last column shows the average MAE metrics over all values of $r$ and shaded cells indicate top-3 good similarity measures whose MAE values are the highest. By convention, we define that preeminent measures (dominant measures) are ones in top-3 lists.

**Table 2.** MAE metric within estimation process

|  | $r$=0.1 | $r$=0.2 | $r$=0.3 | $r$=0.4 | $r$=0.5 | $r$=0.6 | $r$=0.7 | $r$=0.8 | $r$=0.9 | Average (MAE) |
|---|---|---|---|---|---|---|---|---|---|---|
| Jaccard | 0.7465 | 0.7491 | 0.7502 | 0.7543 | 0.7583 | 0.7620 | 0.7717 | 0.7939 | 0.8651 | 0.7723 |
| IJ | 0.7572 | 0.7574 | 0.7578 | 0.7581 | 0.7618 | 0.7658 | 0.7787 | 0.8118 | 0.9135 | 0.7847 |
| Cosine | 0.7532 | 0.7551 | 0.7560 | 0.7593 | 0.7630 | 0.7654 | 0.7736 | 0.7905 | 0.8255 | 0.7713 |
| Pearson | 0.7395 | 0.7462 | 0.7519 | 0.7611 | 0.7734 | 0.7882 | 0.8091 | 0.8435 | 0.8473 | 0.7845 |
| PSS | 0.7452 | 0.7479 | 0.7490 | 0.7529 | 0.7568 | 0.7606 | 0.7708 | 0.7929 | 0.8591 | 0.7706 |
| TA | 0.7518 | 0.7538 | 0.7547 | 0.7581 | 0.7618 | 0.7643 | 0.7726 | 0.7901 | 0.8487 | 0.7729 |
| CosineJ | 0.7459 | 0.7485 | 0.7496 | 0.7537 | 0.7577 | 0.7615 | 0.7712 | 0.7921 | 0.8537 | 0.7704 |
| PearsonJ | 0.7311 | 0.7375 | 0.7427 | 0.7510 | 0.7624 | 0.7766 | 0.7992 | 0.8379 | 0.9173 | 0.7840 |
| PSSJ | 0.7405 | 0.7441 | 0.7456 | 0.7505 | 0.7550 | 0.7605 | 0.7735 | 0.8016 | 0.8718 | 0.7715 |
| TAJ | 0.7449 | 0.7475 | 0.7486 | 0.7527 | 0.7568 | 0.7606 | 0.7704 | 0.7920 | 0.8552 | 0.7699 |
| CosineIJ | 0.7637 | 0.7660 | 0.7668 | 0.7706 | 0.7748 | 0.7767 | 0.7846 | 0.7996 | 0.8515 | 0.7838 |
| PearsonIJ | 0.7580 | 0.7678 | 0.7756 | 0.7864 | 0.8005 | 0.8151 | 0.8321 | 0.8603 | 0.9233 | 0.8132 |
| PSSIJ | 0.7504 | 0.7526 | 0.7534 | 0.7569 | 0.7608 | 0.7638 | 0.7731 | 0.7931 | 0.8569 | 0.7734 |
| TAIJ | 0.7620 | 0.7643 | 0.7651 | 0.7689 | 0.7731 | 0.7751 | 0.7831 | 0.7988 | 0.8525 | 0.7825 |

Top-3 measures according to MAE metric within the estimation process are TAJ, CosineJ, and PSS whose average MAE metrics are 0.7699, 0.7704, and 0.7706, respectively.

Table 3 shows precision metric of all tested measures over all $r$ = 0.1, 0.2, 0.3, 0.4, 0.5, 0.6, 0.7, 0.8, and 0.9 within the recommendation process given precision metric. The last column shows average precision metrics over all values of $r$ and shaded cells indicate top-3 good values.

**Table 3.** Precision metric within recommendation process

|  | $r$=0.1 | $r$=0.2 | $r$=0.3 | $r$=0.4 | $r$=0.5 | $r$=0.6 | $r$=0.7 | $r$=0.8 | $r$=0.9 | Average (Precision) |
|---|---|---|---|---|---|---|---|---|---|---|
| Jaccard | 0.0056 | 0.0105 | 0.0155 | 0.0207 | 0.0261 | 0.0317 | 0.0377 | 0.0438 | 0.0511 | 0.0270 |
| IJ | 0.0062 | 0.0121 | 0.0180 | 0.0240 | 0.0290 | 0.0329 | 0.0321 | 0.0276 | 0.0196 | 0.0224 |
| Cosine | 0.0055 | 0.0104 | 0.0154 | 0.0207 | 0.0262 | 0.0324 | 0.0396 | 0.0508 | 0.0836 | 0.0316 |
| Pearson | 0.0051 | 0.0095 | 0.0141 | 0.0187 | 0.0237 | 0.0291 | 0.0359 | 0.0467 | 0.0803 | 0.0292 |
| PSS | 0.0057 | 0.0106 | 0.0157 | 0.0210 | 0.0266 | 0.0329 | 0.0401 | 0.0512 | 0.0834 | 0.0319 |
| TA | 0.0055 | 0.0104 | 0.0155 | 0.0207 | 0.0263 | 0.0325 | 0.0397 | 0.0509 | 0.0836 | 0.0317 |
| CosineJ | 0.0056 | 0.0105 | 0.0156 | 0.0209 | 0.0265 | 0.0327 | 0.0399 | 0.0510 | 0.0835 | 0.0318 |
| PearsonJ | 0.0052 | 0.0097 | 0.0143 | 0.0190 | 0.0240 | 0.0295 | 0.0362 | 0.0471 | 0.0804 | 0.0295 |
| PSSJ | 0.0057 | 0.0107 | 0.0158 | 0.0212 | 0.0268 | 0.0331 | 0.0403 | 0.0512 | 0.0831 | 0.0320 |
| TAJ | 0.0056 | 0.0106 | 0.0157 | 0.0209 | 0.0265 | 0.0328 | 0.0400 | 0.0511 | 0.0835 | 0.0319 |
| CosineIJ | 0.0055 | 0.0103 | 0.0153 | 0.0205 | 0.0260 | 0.0322 | 0.0394 | 0.0506 | 0.0835 | 0.0315 |
| PearsonIJ | 0.0051 | 0.0095 | 0.0140 | 0.0186 | 0.0236 | 0.0289 | 0.0356 | 0.0465 | 0.0801 | 0.0291 |
| PSSIJ | 0.0056 | 0.0106 | 0.0156 | 0.0209 | 0.0265 | 0.0328 | 0.0399 | 0.0511 | 0.0836 | 0.0318 |
| TAIJ | 0.0055 | 0.0104 | 0.0154 | 0.0206 | 0.0261 | 0.0323 | 0.0395 | 0.0507 | 0.0835 | 0.0316 |

Top-3 measures according to precision metric within recommendation process are PSSJ, PSS, and TAJ, whose average precision metrics are 0.0320, 0.0319, and 0.0319, respectively. From $r$=0.1 to $r$=0.5, IJ is the best measure with precision metric, yet it is no longer preeminent from $r$=0.6 to $r$=0.9; with it is being worse unexpectedly on $r$=0.9. This implies that IJ needs a large amount of training data more than other measures.

Table 4 shows recall metric of all tested measures over all $r$ = 0.1, 0.2, 0.3, 0.4, 0.5, 0.6, 0.7, 0.8, and 0.9 within recommendation process given recall metric. The last column shows average recall metrics over all values of $r$ and shaded cells indicate top-3 good values.

**Table 4.** Recall metric within recommendation process

|  | $r$=0.1 | $r$=0.2 | $r$=0.3 | $r$=0.4 | $r$=0.5 | $r$=0.6 | $r$=0.7 | $r$=0.8 | $r$=0.9 | Average (Recall) |
|---|---|---|---|---|---|---|---|---|---|---|
| Jaccard | 0.9266 | 0.9230 | 0.9221 | 0.9191 | 0.9158 | 0.9155 | 0.9073 | 0.8947 | 0.8496 | 0.9082 |
| IJ | 0.7928 | 0.7514 | 0.6938 | 0.6142 | 0.5099 | 0.3757 | 0.2249 | 0.0940 | 0.0199 | 0.4530 |
| Cosine | 0.9241 | 0.9208 | 0.9211 | 0.9177 | 0.9150 | 0.9147 | 0.9066 | 0.8937 | 0.8021 | 0.9018 |
| Pearson | 0.9439 | 0.9402 | 0.9388 | 0.9359 | 0.9331 | 0.9309 | 0.9190 | 0.8948 | 0.7834 | 0.9133 |
| PSS | 0.9248 | 0.9219 | 0.9215 | 0.9179 | 0.9152 | 0.9143 | 0.9055 | 0.8903 | 0.7936 | 0.9006 |
| TA | 0.9242 | 0.9211 | 0.9211 | 0.9177 | 0.9149 | 0.9145 | 0.9060 | 0.8928 | 0.8005 | 0.9014 |
| CosineJ | 0.9266 | 0.9232 | 0.9223 | 0.9193 | 0.9159 | 0.9153 | 0.9066 | 0.8914 | 0.7970 | 0.9020 |
| PearsonJ | 0.9429 | 0.9440 | 0.9373 | 0.9351 | 0.9323 | 0.9309 | 0.9186 | 0.8948 | 0.7814 | 0.9130 |
| PSSJ | 0.9276 | 0.9239 | 0.9232 | 0.9191 | 0.9160 | 0.9142 | 0.9037 | 0.8844 | 0.7860 | 0.8998 |
| TAJ | 0.9265 | 0.9229 | 0.9224 | 0.9191 | 0.9154 | 0.9149 | 0.9060 | 0.8907 | 0.7957 | 0.9015 |
| CosineIJ | 0.9198 | 0.9160 | 0.9162 | 0.9127 | 0.9100 | 0.9096 | 0.9009 | 0.8886 | 0.8006 | 0.8972 |
| PearsonIJ | 0.9388 | 0.9334 | 0.9308 | 0.9270 | 0.9214 | 0.9181 | 0.9061 | 0.8844 | 0.7787 | 0.9043 |
| PSSIJ | 0.9215 | 0.9189 | 0.9186 | 0.9158 | 0.9132 | 0.9125 | 0.9040 | 0.8905 | 0.7957 | 0.8990 |
| TAIJ | 0.9207 | 0.9165 | 0.9164 | 0.9131 | 0.9099 | 0.9097 | 0.9009 | 0.8878 | 0.7990 | 0.8971 |

Top-3 measures according to recall metric within recommendation process are Pearson, PearsonJ, Jaccard, whose average recall metrics are 0.9133, 0.9130, and 0.9082, respectively.

Finally, F1 metric assembles precision and recall together. The larger F1 is, the better measures are.

$$F1 = \frac{2 * \text{Precision} * \text{Recall}}{\text{Precision} + \text{Recall}} \quad (9)$$

Shortly, MAE is used to evaluate estimation process and F1 is used to evaluate recommendation process. Table 5 which is derived from Tables 2-4 shows the averaged values of both MAE and F1 of all considered measures. The shaded cells indicate the best measures whose values are the highest.

**Table 5.** General MAE and F1 over all measures

|  | MAE | F1 |
|---|---|---|
| Jaccard | 0.7723 | 0.052378 |
| IJ | 0.7847 | 0.042669 |
| Cosine | 0.7713 | 0.061102 |
| Pearson | 0.7845 | 0.056653 |
| PSS | 0.7706 | 0.061638 |
| TA | 0.7729 | 0.061205 |
| CosineJ | 0.7704 | 0.061434 |
| PearsonJ | 0.7840 | 0.057133 |
| PSSJ | 0.7715 | 0.061781 |
| TAJ | 0.7699 | 0.061537 |
| CosineIJ | 0.7838 | 0.060822 |
| PearsonIJ | 0.8132 | 0.056386 |
| PSSIJ | 0.7734 | 0.061510 |
| TAIJ | 0.7825 | 0.060967 |

Top-3 measures according to F1 metric within recommendation process are PSSJ, PSS, and TAJ whose average recall metrics are 0.061781, 0.061638, and 0.061537, respectively.

## 5. Discussion

Based on the experimental results, we find that, two top-3 sets of good measures are {TAJ, CosineJ, PSS}, and {PSSJ, PSS, TAJ}. The preeminent measures are specified as members of the intersection of such three sets which are TAJ and PSS. In fact, it is useful to compare TAJ and PSS but it is impossible to unify metrics MAE and precision / recall together. To overcome this limitation, we can compare them by radar chart, yet some transformations are necessary. Let IMAE be the inverse of normalized MAE. The larger the IMAE is, the better the measures are.

$$\text{IMAE} = 1 - \frac{\text{MAE}}{m} \qquad (10)$$

Note, $m$ is the maximum rating value. Table 6 lists metrics IMAE, precision, and recall of preeminent measures TAJ and PSS.

**Table 6.** Comparison of TAJ and PSS with IMAE, precision, and recall

|     | IMAE   | Precision | Recall  |
|-----|--------|-----------|---------|
| TAJ | 0.8460 | 0.0319    | 0.9015  |
| PSS | 0.8459 | 0.0319    | 0.9006  |

From Table 6, TAJ is better than PSS overall whereas PSS is better than TA with MAE and precision (see Tables 2 and 3). The reason is that TAJ is combined measure of TA and Jaccard. However, it is interesting to learn that TAJ is better than PSSJ in the overall although PSSJ is also a combined measure. The possible reason is that there is over-fitting problem when accuracies among good measures are not so different. Fig. 2 shows radar chart of preeminent measures TAJ and PSS regarding IMAE, precision, and recall.

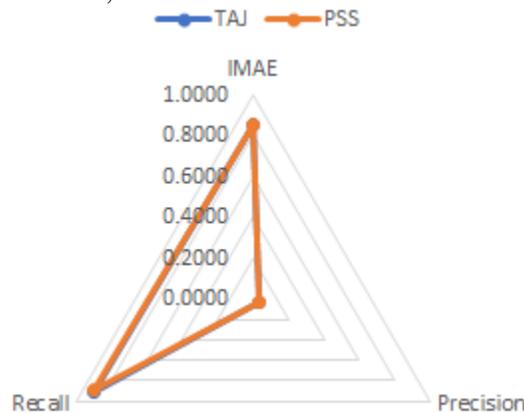

**Figure 2.** Comparison of TAJ and PSS with IMAE, precision, and recall

As seen in Fig. 2, lines of TAJ and PSS are nearly overlapped.

## 6. Conclusions and Future Works

In this work, we combine jaccard with several numerical measures for the purpose of enhacing CF performance. On the whole, it can be seen that from experimental results the combined measures are preeminent ones which takes advantages of data existence and data magnitude. A combined measure has two built-in measures such as Jaccard and a numeric measure. With the built-in Jaccard, the missing values are not ignored and they also contribute to accuracy of recommendation process. With the built-in numeric measure, real number values reflex exactly user favorites. Given numeric measure A, suppose the combined measure of A and Jaccard is called A+. Of course, A+

is often better than A, but it is not asserted that A+ is better than another numeric measure like B, for example. In other words, whether A+ is better than B is dependent on measure A itself. Similarly, IJ is better than traditional Jaccard from $r=0.1$ to $r=0.6$ but PSSIJ is worse than PSSJ from $r=0.1$ to $r=0.6$ with precision metric (see Table 3) because preeminence of the combined measures PSSIJ and PSSJ depends on PSS itself mainly. It is possible that we develop a numeric measure as well as possible and then combine it with Jaccard or variants of Jaccard. However, in practice when A and B are not far different in accuracy, A+ is often better than B. Especially, rating data is always incomplete, in which Jaccard is proved with its good accuracy. Therefore, research on improvement of Jaccard is necessary as aforementioned that Jaccard is an important factor to enhance any numeric measures.

## 6.1. Future Work Proposal

In future work, we aim at developing a new AI based Recommendation System (RS) for Online retails by enhancing collaborative filtering (CF) performance. For this, a comprehensive CF framework with almost ninety similarity measures will be developed to promote CF performance. These similarity measures seek to overcome the recurrent data sparsity in online retail tools by addressing the relationships among the correlated and non-correlated users/items, and this is in a concrete step to build a comprehensive CF framework. The novelty of these similarity measures is embedded in their capability to overcome the recurrent data sparsity in online retail tools by addressing the relationships among the correlated and non-correlated users/items. Moreover, the strength and weaknesses of the proposed similarity measures will be analyzed theoretically and empirically so a comprehensive experimental guide (in terms of effectiveness and efficiency) is developed. This framework will help researchers/Scholars to: (1) find out which similarity measure is better and under which conditions, and (2) identify the measures that could be applied in any circumstances like data sparsity conditions. Furthermore, to overcome the recurrent data sparsity in online retail tools and addressing the relationships among the correlated and non-correlated users/items, new variation of CNN, SVM and MNB algorithms will be developed and integrated in CF similarity measures to promote CF performance and hence improve Recommendation systems to satisfy online retail and their customer's needs. Different programming languages like Python, R and other AI techniques will be used to implement the CF Similarity measures and enhancing the performance of the Recommendation Systems for online retails.

Moreover, to achieve highly sustainable CF performance, word2vector (C-BOW and Skip Gram) will be adapted along advanced variations of KNN, support vector machine (SVM), multi nominal Bayesian (MNB), and Convolutional Neural Network (CNN) will be integrated with the developed similarity measures. Finally, sentiment analysis will be implemented to boost CF rendering. The developed recommendation will be evaluated against the state-of-the-art CF tools on effectiveness and efficiency on several benchmarked datasets, like MovieLens, and Netflix.

### 6.1.1. Future Work Motivations
So, the key motivations can be drawn as follows;
1. It is reported in many literatures that the major factors that makes CF inefficient is the limitation of their similarity measures. Consequently, an extensive investigation of the widely used similarity measures in CF literature along with proposals of new similarity measures will be the priority for our future work. The performance of all measures (including newly proposed ones) will be compared with each other in terms of efficiency and effectiveness. For each measure, efficiency (time and complexity) and effectiveness

(accuracy, precision, recall, F1 measure, and MSE and RMSE) will be examined under both user-based and item-based models. The outcome will be a reliable framework that would contain up to 90 similarity measures, each of which could be selected to satisfy the online retails and customers' needs/conditions.
2. Unlike existing CF similarity measures, we will be developing AI based similarity measures that tackle data sparsity (major obstacle to existing CF tools) and implicitly combining the rated items and non-rated items in calculating similarity of users and/or combining different measures to take advantages of measures' combination. It is expected that combined (hybrid) measures will highly improve accuracy of KNN algorithm, and CF performance as well.

### 6.1.2. Evaluation of the developed CF similarity measures
In this task, the developed CF similarity measures and their combinations will be evaluated and examined using the ML and DL models, on the most broadly used datasets including (but not limited) Movielens (100K, 1000K), Tencent, Epinions, MovieTweetings, DePaulMovie, InCarMusic. The results of this evaluation process will also be compared to existing works on similarity measures.

### 6.1.3. Further metric Evaluation
Six main metrics to assess CF performance will be used for different evaluation processes (estimation and recommendation), namely, mean absolute error (MAE), mean squared error MSE), RMSE, recall, precision and F1 measure. Quality of a CF algorithm like KNN algorithm depends on both estimation and recommendation. This independent evaluation allows us to test measures more objectively, in which estimation process focused on accuracy of KNN algorithm and recommendation process focuses on quality of ANN algorithm. Unlike previous studies, in this work each process is accurately assessed according to its best fit metrics. Run time and complexity of each measure are involved as well to assess techniques' efficiency.